\documentclass{ifacconf}

\usepackage{graphicx}      
\usepackage{natbib}        
\usepackage{mathtools}
\usepackage{subfigure}
\usepackage{cases}
\usepackage{comment}
\usepackage{amsmath}
\usepackage{balance}
\usepackage{amssymb}
\usepackage{amsfonts}
\newcommand{\qedwhite}{\hfill \ensuremath{\Box}}
\newtheorem{myrem}{Remark}
\newtheorem{dfn}{Definition}
\newtheorem{pbm}{Problem}
\newtheorem{exa}{Example}
\newcommand{\rdef}[1]{Definition\,\ref{#1}}
\newcommand{\req}[1]{\eqref{#1}} 
\newcommand{\rpro}[1]{Problem\,\ref{#1}}
\newcommand{\rsec}[1]{Section\,\ref{#1}}
\newcommand{\rfig}[1]{Fig.\,\ref{#1}} 

\begin{document}
\begin{frontmatter}
\title{Control of Timed Discrete Event Systems with Ticked Linear Temporal Logic Constraints} 
\author[first]{Takuma Kinugawa}, 
\author[first]{Kazumune Hashimoto},
\author[first]{Toshimitsu Ushio} 
\address[first]{Graduate school of Engineering Science, Osaka University, Japan.}
\begin{abstract}                
This paper presents a novel method of synthesizing a fragment of a timed discrete event system(TDES),  
introducing a novel linear temporal logic(LTL), called \textit{ticked LTL$_f$}. 
The ticked LTL$_f$ is given as an extension to LTL$_f$, where the semantics is defined over a \textit{finite} execution fragment. Differently from the standard LTL$_f$, the formula is defined as a variant of metric temporal logic formula, where the temporal properties are described by counting the number of \textit{tick} in the fragment of the TDES. 
Moreover, we provide a scheme that encodes the problem into a suitable one that can be solved by an integer linear programming (ILP). 
The effectiveness of the proposed approach is illustrated through a numerical example of a path planning. 


\end{abstract}
\end{frontmatter}

\section{Introduction}
A discrete event system(DES) is useful for the design of a logical high-level controller in many engineering fields such as manufacturing systems, traffic systems, and robotics(\cite{CL2008,CSX2014}).
There are many formalisms of the DES, where its trajectory are represented by a sequence of states and/or events(\cite{SSS2013}).
To model real-time systems, however, we also need information of times when state transitions occur.
Many formalisms including the temporal information in the models of the DES have been proposed(\cite{BHRR1991}). 
\cite{AD1994} proposed a timed automaton that is an extension of an automaton by introducing real-valued variables indicating times elapsed since events occur.
The timed automaton is a dense time model and, as an abstraction of the dense time, a fictitious clock has been introduced(\cite{RS1997,RTTL_what_good}). 
\cite{DESwithTL_1} introduced a timed transition model(TTM) where a discrete time elapse is described by a special event \textit{tick}. 
Moreover, \cite{BW1994} formulated timed discrete event systems (TDES) by a timed transition graph that is a transition graph with state transitions by the event \textit{tick}.

On the other hand, in computer science, the temporal logic(TL) has been developed to specify the trajectories of systems that we verify(\cite{BK2008,Clarke2018}).
For example, in model checking of a non-terminating program, the specification is described by a TL formula and the correctness of the program is verified.  So, the satisfaction relation for the TL formula is defined over infinite trajectories of the verified system. 
Many different temporal logics have been proposed and their expressiveness have been studied. 
Among them, the linear temporal logic(LTL) is often used because it can describe many properties that  specifications often requires such as safety, stability, and progress.
Many approaches to LTL model checking where the specification is described by an LTL formula have been proposed.
A basic idea to solve the LTL model checking is a usage of a tableau and an automata-theoretic approach is widely used.
As alternative approaches, symbolic model checking using binary decision diagrams and bounded model checking using a SAT solver have been developed. 
In the bounded model checking, we search a lasso type trajectory that is a counterexample of the LTL specification. 
\cite{BHJLS2006} proposed efficient encodings for the bounded LTL model checking.


The TL formula has been also leveraged as formal description of a control specification in the DES(\cite{TW1986,JK2006,SU2018}).
Recently, the formal synthesis of control systems has been much attention to (\cite{Belta2017}). 
For example, \cite{KFP2009} describes a high-level specification by an LTL formula and constructed a hybrid controller satisfying the specification.
\cite{WTM2012} proposed receding horizon control for an LTL control specification.
Many path planning problems of mobile robots can be restricted to a finite horizon. 
A controller synthesis problem where a control specification is described by a TL formula, called an LTL$_{f}$ formula, not for infinite trajectories but for finite ones has been proposed (\cite{Zhu2017}).
\cite{Li2019} presented SAT-based LTL$_f$ model checking. 

In verification and control of real-time systems, however, control specifications depend not only on logical constraints but also on the timing at which each event occurs. 
\cite{K1990} proposed metric TL(MTL) for a timed state sequence with a function that assigns the time stamp to each state.
\cite{MN2004} introduced a signal TL that specifies dense-time real-valued signals.  
\cite{RS1997} considered the case where the real-time information is described based on a fictitious clock. 
\cite{O1990} defined real-time TL fro real-time system modeled by the TTM.
\cite{BKD1998} dealt with a synthesis problem of controllers for TDES with a control specification described by an MTL formula. 
\cite{DS2014} proposed an MTL specification interface that translates an MTL specification to a finite timed transition graph used in the synthesis of a timed supervisor. 

In this paper, we provide a novel approach to controller synthesis for TDES, introducing a novel LTL called \textit{ticked LTL}$_f$. 
As with the standard LTL$_f$(\cite{Zhu2017}), the formula will be interpreted over the finite execution fragment, which, as previously mentioned, may be a natural assumption in many path planning problems. 
In contrast to the standard LTL$_f$, the formula in this paper is given as an MTL, where
 temporal properties are described by counting the number of the event \textit{tick} in the fragment of the TDES. 
As we will see later, the problem is formulated to find a suitable (finite) execution fragment of the TDES, such that a given {ticked LTL}$_f$ formula is satisfied. 
Moreover, we provide an encoding scheme such that the problem can be translated into an integer linear programming (ILP). 
Finally, the effectiveness of the proposed approach is illustrated through a numerical example of a path planning. 

The rest of this paper is organized as follows. 
In Section 2, we introduce TDES formulated by \cite{BW1994}. In Section~3, we define syntax and semantics of the ticked LTL$_f$. In Section~4, we provide the problem and an encoding scheme so that it can be translated in to the integer linear programming. 
In Section~5, we apply the proposed approach to a path planning problem of an agent. 
Section~6 concludes the paper.
\section{Timed discrete event system}
In this section, we recall basic definitions of untimed and timed discrete event systems.

\subsection{Discrete event systems}
Let us first define the following untimed discrete event system (DES), which models the untimed behaviors of the transition system:

\begin{dfn}[Untimed DES] \label{def:G_act}
The untimed DES is a tuple $G_{act}=(S_{act}, \Sigma_{act}, \delta_{act}, s_{0,act}, L_{act}, A_{act})$, where
\begin{itemize}
\item   $S_{act}$ is a set of states,
\item   $\Sigma_{act}$ is a set of events,
\item   $\delta_{act} : S_{act} \times \Sigma_{act} \rightarrow S_{act}$ is a transition function,
\item   $s_{0,act}$ is the initial state,
\item   $AP_{act}$ is a set of atomic propositions, and
\item 	  $L_{act} : S_{act} \rightarrow 2^{AP_{act}}$ is a labeling function. \qedwhite
\end{itemize}
\end{dfn}

Next, we incorporate some \textit{timing} properties in $G_{act}$. 
To this end, assume that each event $\sigma \in \Sigma_{act}$ is enabled during a specified time interval $[ l_{\sigma},\ u_{\sigma}]$, where $l_{\sigma} \in \mathbb{N}$, $u_{\sigma} \in  \mathbb{N} \cup \{\infty\}$ with  $l_{\sigma} \leq u_{\sigma}$ are called the lower time and the upper time bound, respectively.
In particular, the event $\sigma$ is called a prospective ({\textit resp.} remote) event if $u_{\sigma} \in \mathbb{N}$ ({\textit resp.} $u_{\sigma} = \infty$). 
Let $\Sigma_{spe}$, $\Sigma_{rem}$ $\subseteq \Sigma_{act}$ be the sets of prospective and remote events, respectively.  Note that $\Sigma_{spe} \cup \Sigma_{rem} = \Sigma_{act}$.
Then, we introduce the following time interval $T_{\sigma}$ for each event $\sigma \in \Sigma_{act}$:
\begin{equation}
\label{T_sigma}
T_{\sigma}= \left\{
\begin{array}{ll}  
\ [ 0,\ u_{\sigma} ]  & \mbox{if } \sigma \in \Sigma_{spe}, \\
\ [ 0,\ l_{\sigma} ] & \mbox{if } \sigma \in \Sigma_{rem}. \\
\end{array} \right.
\end{equation}	 

Moreover, we introduce the \textit{tick} event, which represents the global clock and will be utilized as an additional event to $\Sigma_{act}$. 
Based on the above, a timed DES corresponding to $G_{act}$ is defined as follows (\cite{BW1994}): 

\begin{dfn}[Timed DES]\label{def:G}
A timed DES (TDES) corresponding to $G_{act}$ is a tuple $G=(S, \Sigma, \delta, s_0,AP, L)$ where
\begin{itemize}
\item  $S = S_{act} \times \prod_{\sigma\in \Sigma_{act}} T_{\sigma}$ is a set of states, 
\item   $\Sigma= \Sigma_{act} \bigcup \{\textit{tick}\}$ is a set of events, 
\item    $\delta : S \times \Sigma \rightarrow S$ is a transition function,
\item   $s_0 \in S$ is the initial state, where $s_0 = (s_{0,act}, \{ t_{\sigma,0} | \sigma \in \Sigma_{act} \})$, and $t_{\sigma,0}$ is given by
					\begin{equation}
						t_{\sigma, 0} \coloneqq \left\{
						\begin{array}{ll}  
							\ u_{\sigma}  & \mbox{if } {\sigma} \in \Sigma_{spe} \\
							\ l_{\sigma} & \mbox{if } {\sigma } \in \Sigma_{rem} \\
						\end{array} \right.
					\end{equation}	 
\item  $AP = AP_{act} $ is a set of  atomic propositions,
\item 	$L  : S \rightarrow 2^{AP} $ is a labeling function, where $L(s)=L_{act}(a)$ and $s = (a, \{ t_\sigma | \sigma \in \Sigma_{act} \}) \in S$. \qedwhite
	\end{itemize}
	\end{dfn}

The transition function $\delta$ is a partial function and, for each $s=(a, \{t_{\tau}\ |\ \tau \in \Sigma_{act} \}) \in S$ and $\sigma \in \Sigma$, $\delta(s, \sigma)$ is defined, denoted by $\delta(s, \sigma)!$, if and only if one of the following three conditions holds.
\begin{itemize}
\item[(C1)]  $[\sigma=\textit{tick}] \wedge [\forall \tau \in \Sigma_{spe} ; \delta_{act}(a, \tau)!\Rightarrow t_{\tau}>0]$.
\item[(C2)]  $[\sigma \in \Sigma_{spe}] \wedge [\delta_{act}(a, \sigma)!] \wedge [0 \leq t_{\sigma} \leq u_{\sigma}-l_{\sigma}]$.
\item[(C3)]  $[\sigma \in \Sigma_{rem}] \wedge [\delta_{act}(a, \sigma)!] \wedge [t_{\sigma} =0]$.
\end{itemize}
Note that, by the condition (C1), \textit{tick} is disabled at $s=(a, \{t_{\tau}\ |\ \tau \in \Sigma_{act} \})$ if there exists a prospective event $\tau \in \Sigma_{spe}$ such that $t_{\tau}=0$.
If $\delta(q, \sigma)!$, then $\delta(q, \sigma)=q'=(a', \{t'_{\tau}\ |\ \tau \in \Sigma_{act} \})$ is given as follows.
\begin{enumerate}
\item If $\sigma=tick$, then $a'=a$ and, for each $\tau \in \Sigma_{act}$,
\begin{itemize}
\item if $\tau \in \Sigma_{spe}$, then
\[
t'_{\tau} = \left\{
\begin{array}{ll}
u_{\tau} & \mbox{if } \delta_{act}(a, \tau) \mbox{ is not defined,} \\
t_{\tau}-1 & \mbox{if } [\delta_{act}(a, \tau)!] \wedge [t_{\tau}>0],
\end{array} \right.
\]
\item if $\tau \in \Sigma_{rem}$, then
\[
t'_{\tau} = \left\{
\begin{array}{ll}
l_{\tau} & \mbox{if } \delta_{act}(a, \tau) \mbox{ is not defined,} \\
t_{\tau}-1 & \mbox{if } [\delta_{act}(a, \tau)!] \wedge [t_{\tau}>0], \\
0 & \mbox{if } [\delta_{act}(a, \tau)!] \wedge [t_{\tau}=0] .
\end{array} \right.
\]
\end{itemize}
\item If $\sigma\in \Sigma_{act}$, then $a'=\delta_{act}(a, \sigma)$ and, for each $\tau \in \Sigma_{act}$,
\begin{itemize}
\item if $\tau \not= \sigma$ and $\tau \in \Sigma_{spe}$, then
\[
t'_{\tau} = \left\{
\begin{array}{ll}
u_{\tau} & \mbox{if } \delta_{act}(a', \tau) \mbox{ is not defined,} \\
t_{\tau} & \mbox{if } \delta_{act}(a', \tau)!,
\end{array} \right.
\]
\item if $\tau = \sigma$ and $\tau \in \Sigma_{spe}$, then
\[
t'_{\tau}=u_{\sigma}
\]
\item if $\tau \not= \sigma$ and $\tau \in \Sigma_{rem}$, then
\[
t'_{\tau} = \left\{
\begin{array}{ll}
l_{\tau} & \mbox{if } \delta_{act}(a', \tau) \mbox{ is not defined,} \\
t_{\tau} & \mbox{if } \delta_{act}(a', \tau)!,
\end{array} \right.
\]
\item if $\tau = \sigma$ and $\tau \in \Sigma_{rem}$, then
\[
t'_{\tau}=l_{\sigma}.
\]
\end{itemize}
\end{enumerate}
The informal definition of $\delta$ is omitted in this paper and the reader is referred to \cite{BW1994} for details.

A \textit{finite execution fragment $\pi$ of $G$} is a finite sequence of alternating states and events
\begin{align}\label{execution}
\pi = s(0), e(1), s(1), \ldots, e(H), s(H), 
\end{align}
where $H \in \mathbb{N}_{> 0}$, 
$s(k) \in S$, $\forall k\in\{0,...,H\}$ and $s(0) = s_0$, $(s({k-1}), e({k}), s(k)) \in \delta$, $\forall k\in \{1, ..., H\}$. 
Here, $H$ is called the \textit{length} or \textit{horizon} of $\pi$. 
Moreover, the corresponding sequence of states 
\begin{align}
s(0), s(1), \ldots, s(H) 
\end{align}
is called a \textit{trajectory} of $G$. 
For given \req{execution} and $k\in\{0, ..., H\}$, let $\pi(k) = s(k)$, and 
\begin{align}
\pi(k...) = s(k), e({k+1}), s({k+1}), \ldots, e({H}), s({H}), \notag 
\end{align}
i.e., $\pi(k...)$ denotes the $k$-th suffix of $\pi$. Moreover, for given $k, j\in\{0, ..., H\}$ with $k\leq j$, let $\pi(k...j)$ be the partial suffix given by 
\begin{align}
\pi(k...j) = s(k), e({k+1}), s({k+1}), \ldots, e({j}), s({j}). \notag 
\end{align}
Moreover, for given \req{execution} and $k, j\in \{0, ..., H\}$ with $k \leq j \leq H$, let $count_\pi(k,j)$ denote the number of the event \textit{tick} occurred in $\pi(k...j)$.
For example, if $\pi=a,tick,a,\sigma,b,tick,a$ with $AP=\{a,b\}$ and $\Sigma = \{\sigma\}\cup \{tick\}$, we have $count_\pi(0,3) = 2$, $count_\pi (1, 3) = 1$ since $\pi(0...3) = a,tick,a,\sigma,b,tick,a$ and $\pi(1...3) = a,\sigma,b,tick,a$. Note that we have $count_\pi(k, k)=0$, $\forall k \in \{ 0, \ldots, H \}$, since $\pi(k...k) = s(k)$ and so no events occur in $\pi(k...k)$.


\section{Ticked linear temporal logic}
We now introduce a novel temporal logic called LTL$_f$.
As will be seen below, this formula is interpreted over a finite execution fragment \req{execution}, and provides an extension to the LTL$_f$ formula (\cite{Zhu2017}), in the sense that we incorporate some timing properties via \textit{tick} events. 
First, we define its syntax as follows: 

\begin{dfn}[Syntax of ticked LTL$_f$]\label{dfn:syntax}
A ticked LTL$_f$ formula over a set of atomic propositions $AP$ is recursively defined according to the following grammar: 
\begin{align}
\phi \coloneqq True \ |\ ap\ |\ \lnot \phi\ |\ \phi_1\land \phi_2\ |\ \phi_1U_{[m,n]} \phi_2 ,
\end{align}
where $ap \in AP$, $m$ and $n$ are nonnegative integers with $m \leq n$. \qedwhite
\end{dfn}
	
Note that we do not include the operator $\bigcirc$ (next) in the syntax, which will not be utilized to express the specification in this paper. Additional boolean operators are defined as 
\begin{align}
\phi_1 \lor \phi_2 &\coloneqq \lnot(\lnot \phi_1 \land \lnot \phi_2),\ \phi_1 \rightarrow \phi_2 \coloneqq \lnot \phi_1 \lor \phi_2 \\ 
\phi_1 \leftrightarrow \phi_2 &\coloneqq (\phi_1 \rightarrow \phi_2)\land(\phi_2 \rightarrow \phi_1). 
\end{align} 
Moreover, other temporal operators, such as $\Diamond_{[m,n]}$ (future) and $\Box_{[m,n]}$ (globally) are defined by
\begin{align}\label{futureglobal}
\Diamond_{[m,n]} \phi \coloneqq True U_{[m,n]} \phi, \ \Box_{[m,n]} \phi \coloneqq \lnot \Diamond_{[m,n]} \lnot \phi. 
\end{align}
	
Its semantics is defined over a finite execution fragment in \req{execution} and is formally given as follows: 
\begin{dfn}[Semantics of ticked LTL$_f$]
Given a finite execution fragment $\pi = s(0), e(1), s(1), \ldots, e(H), s(H)$, the satisfaction of the ticked LTL$_f$ formula $\phi$ for the $k$-th suffix of $\pi$ ($0\leq k \leq H$), denoted as $\pi (k...) \models \phi$, is defined recursively as follows:  
\begin{itemize}
\item 	$\pi(k...) \models True$,
\item 	$\pi(k...) \models ap$ if and only if $ap \in L(\pi(k))$,
\item 	$\pi(k...) \models \lnot \phi$ if and only if $\pi(k...) \not\models \phi$,
\item 	$\pi(k...) \models \phi_1\land \phi_2$ if and only if $\pi(k...) \models \phi_1 \land \pi(k...)\models \phi_2$,
\item 	$\pi(k...) \models \phi_1U_{[m,n]} \phi_2$ if and only if there exist $j \in [k ,H]$ such that $m \leq count_\pi(k, j) \leq n$, $\pi(j...) \models \phi_2$ and $\pi(i...) \models \phi_1$, $\forall i \in [k,j-1]$. \qedwhite
	\end{itemize}
	\end{dfn}
Intuitively, the formula $\phi_1 U_{[m,n]}  \phi_2$ indicates that, $\phi_1$ holds true until $\phi_2$ holds true during the interval that the number of ticked events is between $m$ and $n$. 
We denote by $\pi \models \phi$ if and only if $\pi (0...) \models \phi$.


\textit{(Example):} Consider a finite execution fragment:  
\begin{align}
\pi=a,tick,a,\sigma,b,tick,a. 
\end{align}
Also, consider a ticked LTL$_f$ formula $\phi=a U_{[1,3]} b$, with $AP=\{a,b\}$ and $\Sigma = \{\sigma\}\cup \{tick\}$. Then, $\pi(0...) (= a,tick,a,\sigma,b,tick,a)$ satisfies $\phi$, since $a$ holds true until $b$ holds true while the number of \textit{tick} counted from $\pi(0)$ is $1$, i.e., $count_{\pi}(0, 2) = 1 \in[ 1, 3]$. However, $\pi (1...) (= a,\sigma,b,tick,a)$ does \textit{not} satisfy $\phi$, since $b$ holds true while the number of \textit{tick} counted from $\pi(1)$ is $0$, i.e., $count_{\pi}(1, 2) = 0 \notin [1, 3]$. \qedwhite 

\section{Controller Synthesis under LTL$_f$ constraints}

Using the ticked LTL$_f$ introduced in the previous section, we consider the following problem. 
\begin{pbm}
\label{pbm:1}
Given a TDES $G$, a ticked LTL$_f$ formula $\phi$ and a horizon $H>0$, synthesize a finite fragment $\pi$ of $G$ with the horizon $H$, such that $\pi \models \phi$. \qedwhite 
\end{pbm}

To solve \rpro{pbm:1}, we translate a finite trajectory of the TDES $G$, the counting function $count_\pi$, and the ticked LTL$_f$ formula $\phi$ into a set of integer-valued equations 
that can be solved by integer linear programming (ILP). Details for the encodings are described below. 


\subsection{Encoding the trajectory of $G$}
To encode the trajectory of $G$, we denote by $A \in \{0, 1\}^{N \times N}$ with $N=|S|$ the \textit{adjacency matrix} of the graph in accordance with $G$, i.e., letting $S = \{s_1, ..., s_{N}\}$, we have $A_{i,j} = 1$ (the $(i,j)$-component of $A$ is $1$) if and only if there exists $\sigma \in \Sigma$ such that $s_j \in \delta(s_i, \sigma)$, and $0$ otherwise. Moreover, we introduce $H+1$ binary vectors $w (k) \in \{0, 1\}^{N}$, $k\in \{0, ..., H\}$ to represent the state of $G$ at $k$, where, for each $k \in \{0, ..., H\}$, the vector $w(k)$ includes only one non-zero component. That is, if $\pi$ is given by \req{execution}, we have $w_i (k) = 1$ (the $i$-th component of $w(k)$ is $1$) if and only if $s(k) = s_i$, and $0$ otherwise. The trajectory of the states can be then encoded as follows: 
\begin{equation}\label{w}
w(k+1) \leq A^\mathsf{T} w(k), \ {1}^\mathsf{T} _N w(k) =1, 
\end{equation}
where ${1}_N$ is the $N$-dimensional vector that contains $1$ for all components.

\subsection{Encoding the counting function}
Let $c (k, j) \in \mathbb{N}$ for $k, j \in \{0, ... H\}$ with $k\leq j$ be integer variables that represent the number of \textit{tick} events occurred in $\pi(k...j)$, i.e., $c(k,j) = m$ if and only if $count_\pi(k,j) = m$. This variable can be encoded by the ILP constraints as follows. First, we introduce $H$ binary variables $z_e (k) \in \{0, 1\}$, for $k\in\{1, ..., H\}$ in order to represent the occurrence of \textit{tick} in the sequence of events, i.e., if $\pi$ is given by \req{execution}, we have $z_e(k) = 1$ if and only if $e(k) = tick$. Using $z_e (k)$, $k \in \{0, ..., H \}$, $c(k, j)$ is then given by
\begin{align}\label{ckj}
c (k, j) =   \sum^j _{i = k+1} z_e (i) 
\end{align}
for $k, j \in \{0, ..., H\}$ with $k < j$, and $c (k, k) = 0$, $\forall k\in\{0, ..., H\}$. 
The variables $z_e (k)$, $k\in\{1, ..., H\}$ can be encoded as follows. First, let $\alpha \in \{0, 1\}^{N}$ be a binary vector, such that $\alpha_i = 1$ (the $i$-th component of $\alpha$ is $1$) if and only if $\delta (s_i, tick) !$ (i.e., $s_i$ can transition through the event \textit{tick}). Moreover, let $\beta \in \{0, 1\}^{N}$ be a binary vector, such that $\beta_i = 1$ (the $i$-th component of $\beta$ is $1$) if and only if there exists $s_{j} \in S$, such that $s_i = \delta (s_{j}, tick)$ (i.e., there exists a state that can transition to $s_i$ through the event \textit{tick}).
Then, $z_e (k) = 1$ if and only if 
\begin{align}
\alpha^\mathsf{T} w (k-1) = 1 \wedge \beta^\mathsf{T} w (k) =1. \notag 
\end{align}
Thus, $z_e (k)$ is expressed as follows: 
\begin{align}
z_e (k) &\leq \alpha^\mathsf{T} w (k-1), \label{zek1}\\ 
z_e (k) &\leq \beta^\mathsf{T} w (k) \\ 
z_e (k) & \geq -1 + \alpha^\mathsf{T} w (k-1) + \beta^\mathsf{T} w (k). \label{zek2}
\end{align}

\subsection{Encoding the ticked LTL$_f$ formula}\label{encLTL}
We introduce $H+1$ binary variables $z_{\phi}(k) \in \{0,1\}$ for $k \in \{0,1,...,H\}$, such that $z_{\phi}(k)=1$ if and only if $\pi (k...)$ satisfies $\phi$. The encodings for the ticked LTL$_f$ formula $\phi$ can be recursively given as follows: 

\textit{(atomic proposition):} Let $\phi = ap \in AP$ and $v \in \{0, 1\}^N$ be a binary vector, such that $v_i = 1$ (the $i$-th component of $v$ is $1$) if and only if $ap \in L(s_i)$. Then, the satisfaction of the formula $\phi$ can be encoded as follows: 
\begin{align}
v^\mathsf{T} w(k) &\geq z_\phi (k),\\
v^\mathsf{T} w(k) &< z_\phi (k)+1. 
\end{align}

\textit{(negation):} Let $\phi = \neg \psi$. Then, the satisfaction of $\phi$ can be encoded as 
\begin{align}
z_\phi (k) = 1 - z_\psi (k). 
\end{align}
\textit{(conjunction):} Let $\phi = \bigwedge^L _{\ell=1} \psi_\ell$. Then, 
\begin{align}
\label{translate_con}
z_\phi (k) &\leq z_{\psi_\ell} (k),\ \forall \ell \in \{1, ..., L\}, \notag \\ 
z_\phi (k) & \geq 1-L + \sum_{\ell=1}^L z_{\psi_\ell} (k). 
\end{align}
\textit{(disjunction):} Let $\phi = \bigvee^L _{\ell=1} \psi_\ell$. 
Then, 
\begin{align}
z_\phi (k) &\geq z_{\psi_\ell} (k),\ \forall \ell \in \{1, ..., L\}, \notag \\ 
z_\phi (k) & \leq \sum_{\ell=1}^L z_{\psi_\ell} (k). 
\end{align}
With rough notation, boolean operators are used for binary variables. For example, when we consider $\phi = \bigwedge^L _{\ell=1} \psi_\ell$, we write $z_\phi = \bigwedge^L _{\ell=1} z_{\psi_\ell}$ instead of \req{translate_con}. Then, we describe the translation of temporal operator \textit{until} with this notation.

\textit{(until):} 
Let $\phi = \psi_1 U_{[m,n]} \psi_2$. We introduce binary variables $\underline{z}_c (k, j), \overline{z}_c (k, j) \in \{0, 1\}$, for $k, j \in \{0, ... H\}$ with $k\leq j$,
such that $\underline{z}_c (k,j) = 1$ (resp. $\overline{z}_c (k, j) = 1$) if and only if $m \leq c (k, j)$ (resp. $c(k,j) \leq n$). That is, $c(k, j)$ is encoded as 
\begin{align}
m - M \leq c (k, j) - M \underline{z}_c(k,j) < m \\ 
n  < c (k, j) + M \overline{z}_c(k,j) \leq n+M, 
\end{align}
where $M$ is a sufficiently large number satisfying $M >n$. 
Then, the satisfaction of $\phi$ can be encoded as 
\begin{align}\label{until1}
z_\phi (k) = \bigvee^{H} _{j=k} z_\phi (k, j), 
\end{align}
where 
\begin{align}
z_\phi (k, j) = \overline{z}_c (k,j) \wedge  \underline{z}_c (k,j) \wedge z_{\psi_2} (j) \wedge \left(\bigwedge^{j-1} _{\ell=k} z_{\psi_1} (\ell) \right).  \notag 
\end{align}

The encodings for $\Diamond_{[m,n]}$ and $\Box_{[m,n]}$ can be easily done from the relation \req{futureglobal} and are thus omitted for brevity. 

\subsection{Overall problem}
Based on the above encodings, we can formulate the ILP as follows: 
\begin{numcases}
{{\rm find}:} 
w(k), z_\phi (k), k\in \{0, ..., H\}, \\ 
z_e(k), k\in \{1, ..., H\}, \\ 
c(k, j) , k, j \in \{0, ..., H\}, k \leq j, 
\end{numcases}
subject to the following constraints: 
\begin{align}
\req{w}-\req{ckj},\ ILP(\phi),\ z_\phi (0) = 1, 
\end{align}
where $ILP(\phi)$ is the ILP constraints for ticked LTL$_f$ formula $\phi$ generated from the procedure described in \rsec{encLTL}. The above problem can be solved by several off-the-shelf tools, such as Gurobi (available: https:\slash\slash www.gurobi.com), z3 (\cite{MB2008}), and so on. 


\section{Application to path planning}
In this section, we demonstrate the effectiveness of the proposed approach through a numerical simulation of a path planning. 
\subsection{Setting of TDES}
\begin{figure}
	\begin{center}
		\includegraphics[width=6cm]{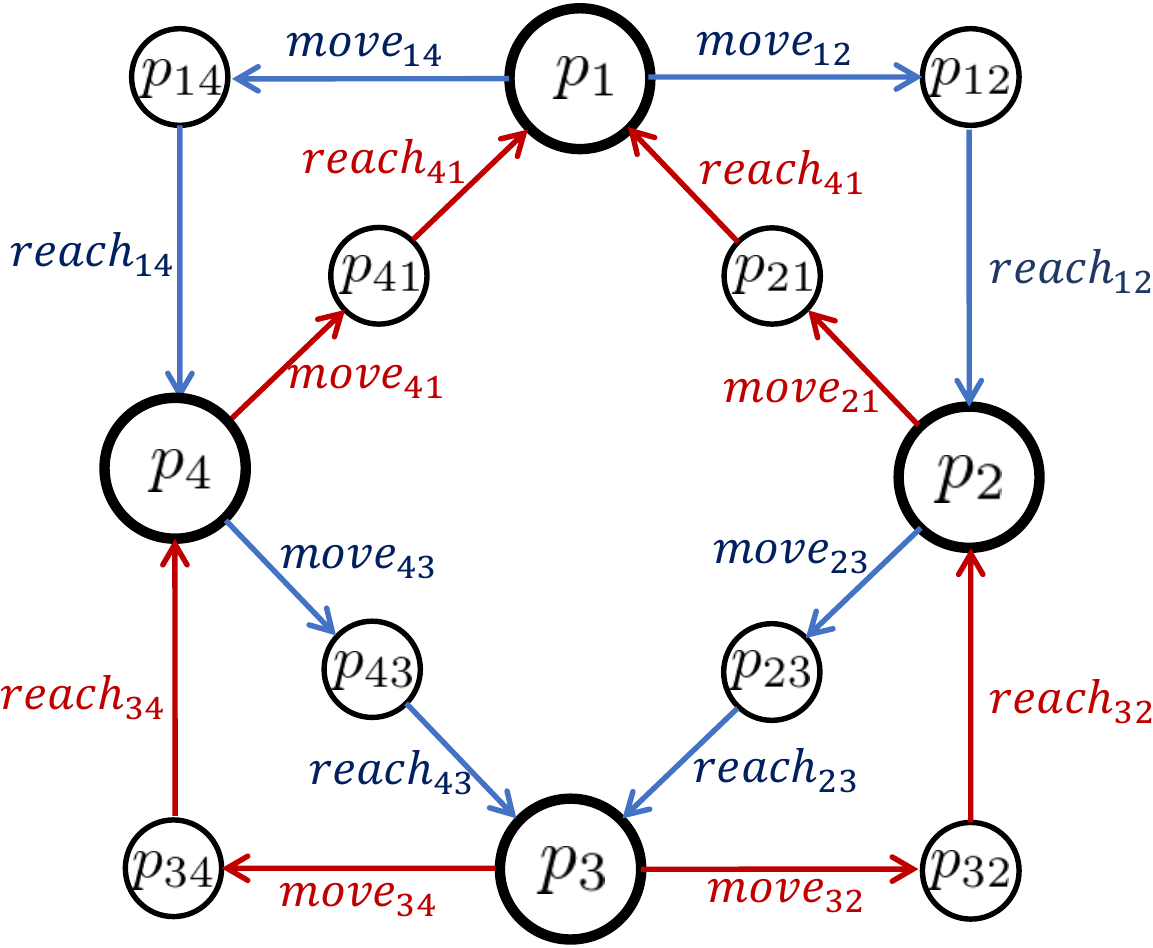}
			\caption{The untimed transition system $G_{act}$ considered in the simulation example. }
			\vspace{0.5cm}
			\label{gactfig}
		\end{center}
\end{figure}

The agent (e.g., robot, drone, \textit{etc}) is first represented by the untimed transition system $G_{act}$, as shown in \rfig{gactfig}. In the figure, each node represents the state of the agent, and each edge represents the transition among them. 
More specifically, if the state of the agent is $p_{i}$ ($i\in\{1,\ldots,4\}$), it means that \textit{the agent is in the location $p_i$}. 
Moreover, if the state is $p_{i j}$, it means that \textit{the agent is on the way from $p_i$ to $p_j$}. The symbols $move_{i j}$ and $reach_{ij}$, $i, j \in \{1, \ldots, 4\}$ represent the events that are associated to the edges. 
More specifically, the event $move_{i j}$ indicates that the agent decides to move from $p_i$ to $p_j$, and the event $reach_{i j}$ indicates that the agent reaches $p_j$. 
The set of atomic propositions is given by $AP _{act} =\{ap_1, ap_2, ap_3, ap_4\}$, and the labeling function is $L_{act}(p_i) = ap_i$, $\forall i\in\{1, ... 4\}$. The initial state is $s_{0, act} = p_1$. 

The time interval $T_\sigma$ is then defined as follows: if $reach_{i,j}$ for $i, j\in\{1, \ldots, 4\}$ is defined in \rfig{gactfig}, $T_{reach_{ij}}$ is given by 
\begin{align}
&T_{reach_{12}}=[2, \infty], T_{reach_{14}}=[1, \infty], T_{reach_{21}}=[3, \infty],\notag \\
&T_{reach_{23}}=[2, \infty], T_{reach_{32}}=[3, \infty], T_{reach_{34}}=[2, \infty],\notag \\
&T_{reach_{41}}=[2, \infty], T_{reach_{43}}=[1, \infty]. \label{tsigma1}
\end{align}

For example, $T_{reach_{12}}=[2, \infty]$ implies that, if the state of the agent is $p_{12}$ (i.e., it is on the way from $p_1$ to $p_2$), the event $reach_{12}$ can occur at any time after $2$ ticks. In other words, the agent requires at least $2$ ticks to reach from $p_i$ to $p_j$. 
On the other hand, if $move_{ij}$ for $i, j\in\{1, \ldots, 4\}$ is defined in \rfig{gactfig}, $T_{move_{ij}}$ is then given by 
\begin{align}\label{tsigma2}
T_{move_{ij}}=[0, \infty], 
\end{align}
\req{tsigma2} indicates that, if the state of the agent is $s_i$, the event $move_{ij}$ can occur at any time (i.e., with any number of ticks).  
Based on the above, the corresponding timed transition system $G$ is constructed according to \rdef{def:G}.

\subsection{Simulation results}
\begin{figure}[t]
   \centering
    \subfigure[Partial fragment of $\pi_1$ until the number of \textit{tick} event is $1$.]{
      {\includegraphics[width=8.5cm]{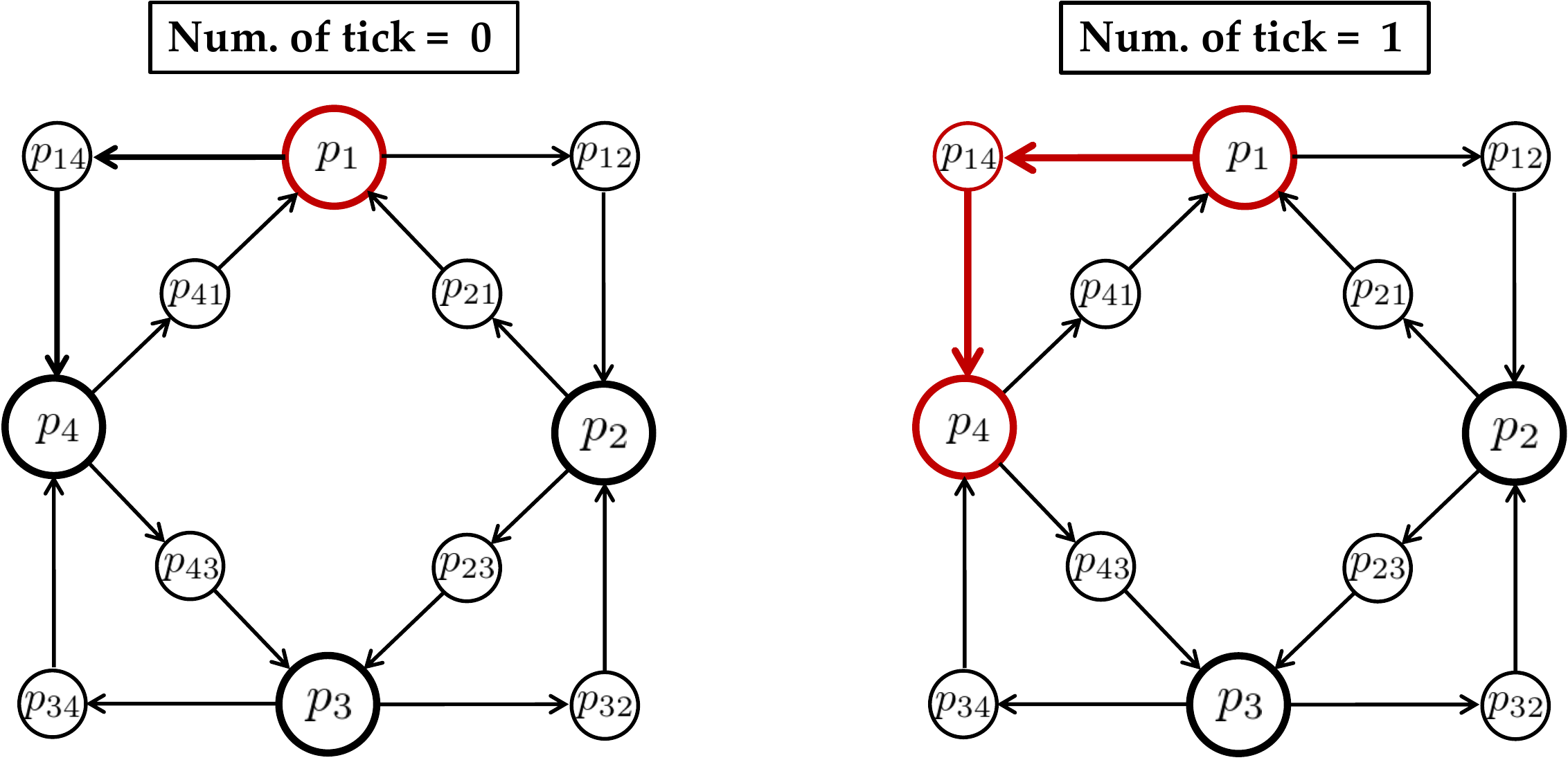}\vspace{0cm}} \label{result11}}
    \subfigure[Partial fragment of $\pi_1$ until the number of \textit{tick} event is $5$.]{
      {\includegraphics[width=8.5cm]{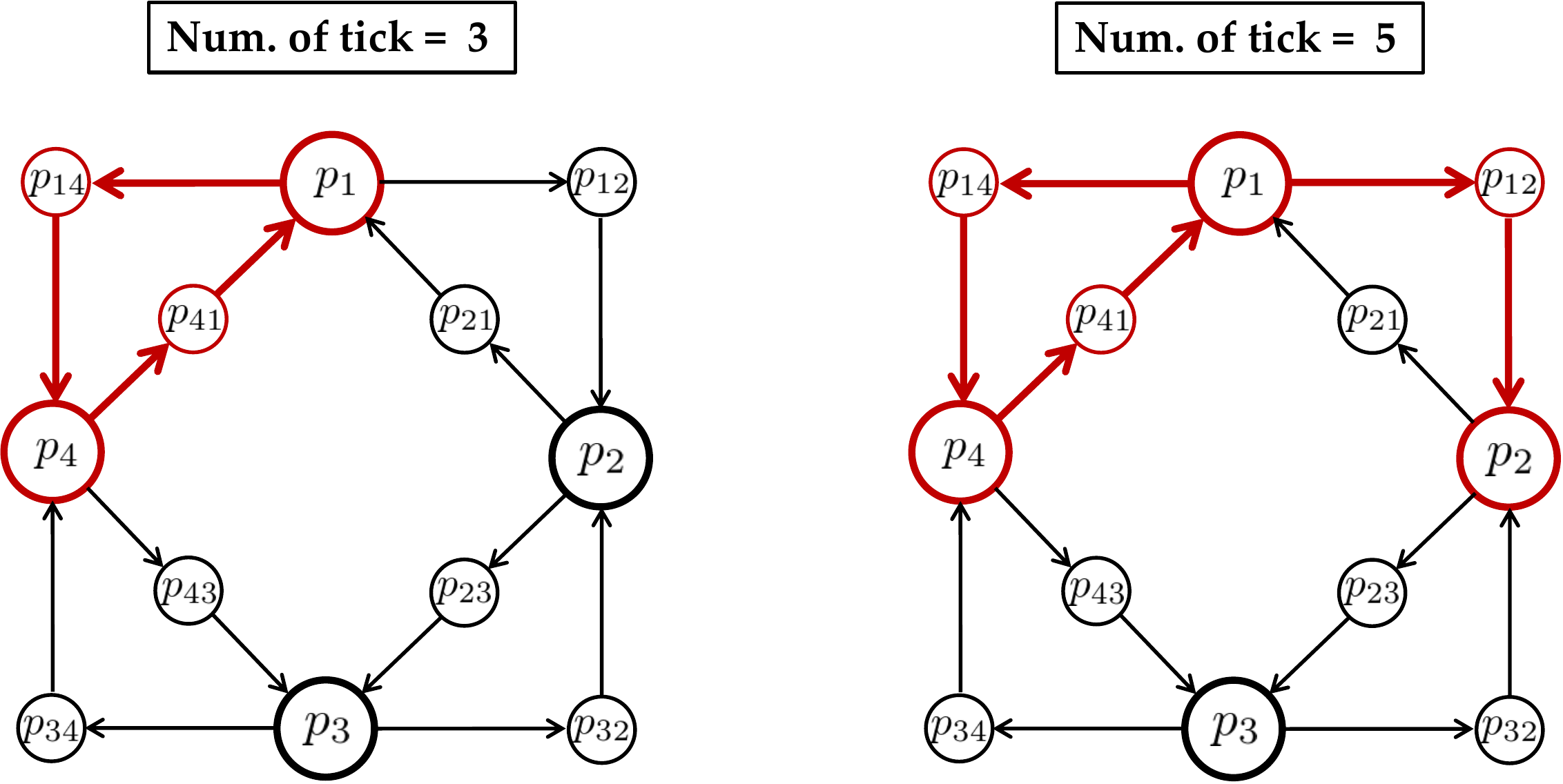}}\label{result12} }
    \caption{Resulting execution fragment $\pi_1$ by solving the ILP. In the figure, red nodes and edges represent the path that the agent traverses according to $\pi_1$.} \label{result1}
    \vspace{0.5cm}
\end{figure}

We first consider the following specification: $\phi_1=\Diamond_{[1,5]}ap_2 \land \Diamond_{[1,5]}ap_4$. That is, starting from the initial position (i.e., $s_{0, act} = p_1$), the agent must reach $p_2$ and $p_4$ while the number of the event \textit{tick} is between $1$ and $5$. The corresponding ILP is solved with different selections of $H$, in order to find the execution fragment satisfying $\phi_1$. 
Specifically, starting from $H=5$, we solve the corresponding ILP and we increment the horizon until the execution fragment satisfying $\phi_1$ has been found. 
The execution fragment was found with $H = 11$ and is illustrated in \rfig{result1}. The figure shows that $\Diamond_{[1,5]}ap_4$ is satisfied with the total number of \textit{tick} given by $1$ (right figure of \rfig{result11}), and $\Diamond_{[1,5]}ap_2$ is satisfied with the total number of \textit{tick} given by $5$ (right figure of \rfig{result12}). The resulting fragment is concretely given by 
\begin{align}
\pi_1 = &\ p_1, move_{14}, p_{14}, tick, p_{14}, reach_{14}, p_4,... \notag \\ 
       &\ move_{41}, p_{41}, tick, p_{41}, tick, p_{41}, reach_{41}, p_1,... \notag \\ 
       &\ move_{12}, p_{12}, tick, p_{12}, tick, p_{12}, reach_{12}, p_2. \label{fragex}
\end{align}
Therefore, the resulting execution fragment is shown to satisfy $\phi_1$. 
\req{fragex} implies that the agent aims to satisfy $ap_4$ and then satisfy $ap_2$. 
Alternatively, the agent \textit{might} instead aim to satisfy $ap_2$ and then $ap_4$. 
However, from \req{tsigma1}, this would then require at least 
$2+3+1 = 6$ ticks to reach $p_4$, which means that the formula $\Diamond_{[1,5]}ap_4$ does \textit{not} hold. That is, if the fragment \textit{were} generated such that the agent aims to satisfy $ap_2$ and then $ap_4$, it would then violate $\phi_1$. Hence, it is shown that the ILP could appropriately select the fragment, such that the agent could satisfy the desired specification. 

\begin{figure}[t]
   \centering
    \subfigure[Partial fragment of $\pi_2$ until the number of \textit{tick} event is $1$.]{
      {\includegraphics[width=8.5cm]{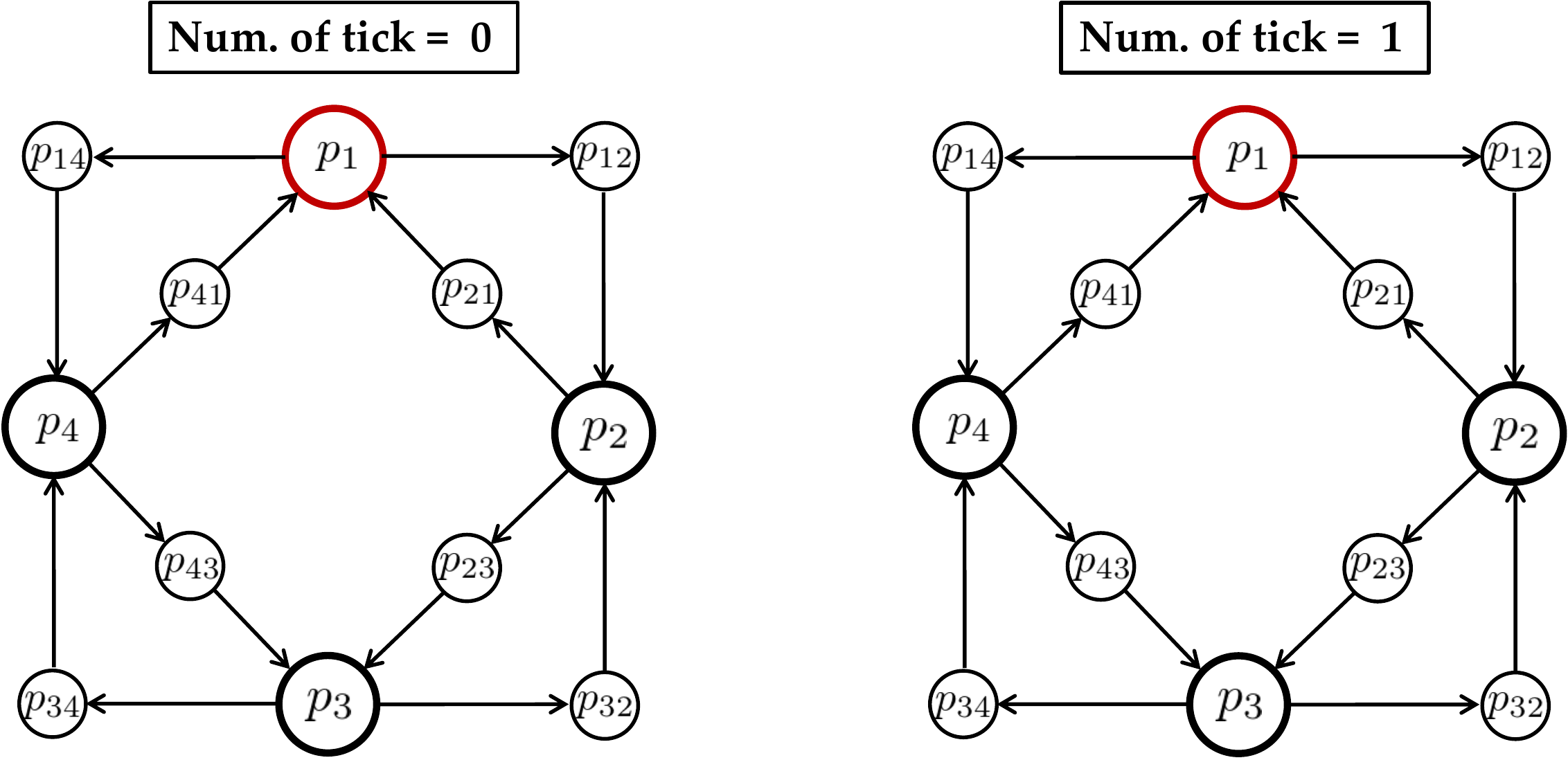}\vspace{0cm}} \label{result21}}
    \subfigure[Partial fragment of $\pi_2$ until the number of \textit{tick} event is $3$.]{
      {\includegraphics[width=8.5cm]{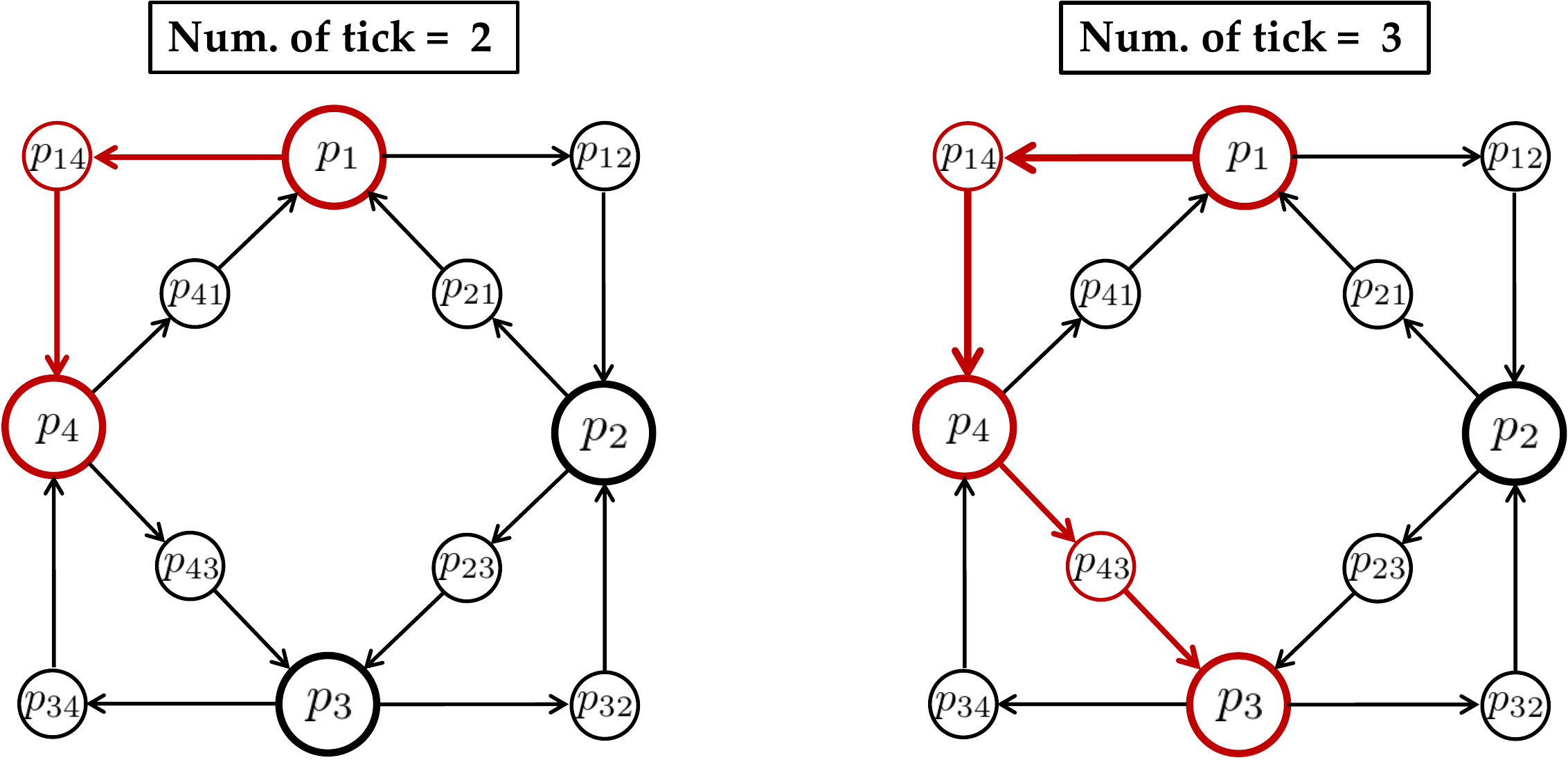}}\label{result22} }
    \caption{Resulting fragment $\pi_2$ by solving the ILP. In the figure, red nodes and edges represent the path that the agent traverses according to $\pi_2$.} \label{result2}
    \vspace{0.5cm}
\end{figure}

As another example, we consider $\phi_2=(\lnot ap_2) U_{[3,5]} ap_3$, which indicates that the agent must avoid $p_2$ until the agent reaches $p_3$ with the number of \textit{tick} being from $3$ to $5$. 
The execution fragment satisfying $\phi_2$ is found with $H = 10$ and the result is shown in \rfig{result2}. 
The figure shows that the agent reaches $p_3$ while avoiding $p_2$ with total number of \textit{tick} given by $3$ (Fig.\ref{result22}). The resulting fragment is concretely given by
\begin{align}
\pi_2 = &\ p_1, tick, p_1, move_{14}, p_{14}, tick, p_{14}, reach_{14}, p_4,... \notag \\
		  &\ move{43}, p_{43}, tick, p_{43}, reach_{43}, p_3, move_{32}, p_{32}, ... \notag \\
		  &\ tick, p_{32}.  \label{fragex2}
\end{align}
Therefore, the it is shown that the agent satisfies the formula $\phi_2$. 

\section{Conclusion and future work}
In this paper, we considered a TDES proposed by Brandin and Wonham, where the elapse of time is described by an event $tick$, and propose ticked LTL$_f$ that describes real-time constraints based on the occurrence of $tick$ in the TDES.
To find the solution of \rpro{pbm:1} we provide an approach to encode \rpro{pbm:1} into ILP. Then, we illustrate the effectiveness of the proposed approach through a numerical example. 

Note that this paper deals with the problem of finding a \textit{feasible} execution fragment of TDES, such that the ticked LTL$_f$ is satisfied. Hence, future work involves finding an \textit{optimal} execution fragment, such that a certain cost function is minimized while the ticked LTL$_f$ is satisfied. The authors believe that this will be achieved by providing an encoding scheme so that the problem can be solved by a MAX-SAT solver. 

\section*{Acknowledgement}
The authors are supported by ERATO HASUO Metamathematics for Systems Design Project (No. JPMJER1603), JST.
\balance 

\end{document}